\documentclass[twocolumn,showpacs,superscriptaddress,amsmath,amssymb,aps,pra]{revtex4-1}
\usepackage{graphicx}
\usepackage{CJKutf8}
\usepackage{dcolumn}
\usepackage{amsmath,bm}
\usepackage{epstopdf}
\usepackage[mathlines]{lineno}
\usepackage{color}
\usepackage[colorlinks=true,linkcolor=blue,citecolor=magenta,urlcolor=cyan]{hyperref}

\begin{document}
\title{Corner states are not topologically protected}
\author{Yi-Dong Wu}
\email{wuyidong@ysu.edu.cn}
\affiliation{Key Laboratory for Microstructural Material Physics of Hebei Province, School of Science, Yanshan University, Qinhuangdao 066004, China}

\begin{abstract}
Recently there is a surge of interests in the so-called topologically protected corner states in 2D and 3D systems. Such systems are considered as high order topological insulators. Wannier centers are used as topological invariants to characterize bulk systems. The existence of corner states is considered as a reflection of the topological non-triviality of bulk energy bands. We demonstrate that the Wannier centers are not topological invariants by showing they depend on the choices of unit cells. The same bulk system can be considered as both topological and trivial with two equally possible types of unit cells. We show the existence of corner states only reflects different choices of boundaries of the same bulk system. The corner states disappears in the so-called topological state if we choose a different boundary with the same symmetry as the original one. On the other hand, equally robust corner states can be realized in the so-called trivial state.
\end{abstract}
\date{\today}
\maketitle

The existence of localized states in 1D system has been studied intensely. The edge state in 1D systems can be considered as the prototype of the corner state in higher dimensional systems. For example, in the 1D SSH model Zak phase $\phi_{Zak}$ is considered as topological invariant\cite{PhysRevB.84.195452,PhysRevLett.110.180403,PhysRevB.94.125119,PhysRevB.89.085111,atala2013direct}. The bulk state of system is claimed to be topological when $\phi_{Zak}=\pi$ and trivial when $\phi_{Zak}=0$. The presence or absence of the zero-energy edge states is considered as indication of the topological property of the bulk bands. It is claimed the topological property of this system is protected by the chiral symmetry\cite{PhysRevB.84.195452}.

The corner states in 2D and 3D systems have been proposed and observed recently\cite{Benalcazar61,PhysRevB.96.245115,imhof2018topolectrical-circuit,serragarcia2018observation,zhang2019second-order,PhysRevLett.123.073601,peterson2018a,PhysRevLett.120.026801,PhysRevB.98.045125,
kempkes2019robust,ni2019observation,mittal2019photonic,hassan2019corner,PhysRevLett.122.244301,PhysRevB.99.245151,PhysRevB.100.075120,PhysRevB.100.075418,PhysRevB.100.205109}. The corner states, similar to the edge states in 1D systems, are also claimed to be topologically protected. The polarization of occupied bands\cite{PhysRevLett.120.026801} and polarization of the Wannier bands\cite{Benalcazar61} are used as topological invariant to characterize the bulk systems. When the polarization take certain values the system is considered as topological, otherwise it is considered as trivial.

Here we first show neither of Zak phase in 1D systems nor the polarizations in 2D systems can be considered as bulk topological invariant. For this purpose, we stress some characteristics of topological invariant of bulk bands. First, as a property of energy bands, a bulk topological invariant must be defined without referring to the boundary of the system. Since the Bloch wave can only be defined in a system with infinite boundary condition or periodic boundary condition, i.e. a system without terminations, this conclusion is obvious.

Second, since one bulk system must be in one definite topological state, i.e. it is either topological or trivial, the topological invariants calculated from the same bulk bands must indicate the same topological state of the system. For example, we can choose different types of unit cells in calculating the topological invariant. If the so-called topological invariant indicates the system is topological with one type of unit cell and indicates the system is trivial with another type, then it is not a topological invariant.

We now show Zak phase and the polarizations depend on the choices of unit cells. Zak phase and the polarizations correspond to the centers of Wannier functions\cite{Benalcazar61,PhysRevB.96.245115}. The unit-cell-dependence of Wannier centers come from the fact that the positions of the orbits in a tight-binding model are not uniquely defined. If we use the Berry connection to calculate the Wannier centers the positions of the orbits depend on the choice of unit cells.

In 1D systems Zak phase $\phi_{Zak}$ and the  Wannier center $x_w$ are related by $\phi_{Zak}=x_w2\pi$. It can be easily shown
\begin{equation}\label{zak}
\begin{split}
x_w &=\frac{1}{2\pi}i \int_0^{2\pi}\langle u (k)|\frac{\partial}{\partial k}|u (k)\rangle dk\\
             &=\sum\limits_{n}n \langle W(n)|W(n)\rangle,
\end{split}
\end{equation}
where $|W(n)\rangle$ is the Wannier function. For given lattice $n$, $|W(n)\rangle$ is a column vector and the elements of $|W(n)\rangle$ correspond to orbits in the unit cell denoted by $n$. $\langle W(n)|W(n)\rangle$ is the probability that the electron in this unit cell. All the orbits in $n$th unit cell are assumed to be located at $n$ in calculating $x_w$. So the Wannier center obtained from the Zak phase is not actual the average position of electron for a given Wannier function.

\begin{figure}
  \includegraphics[width=8cm]{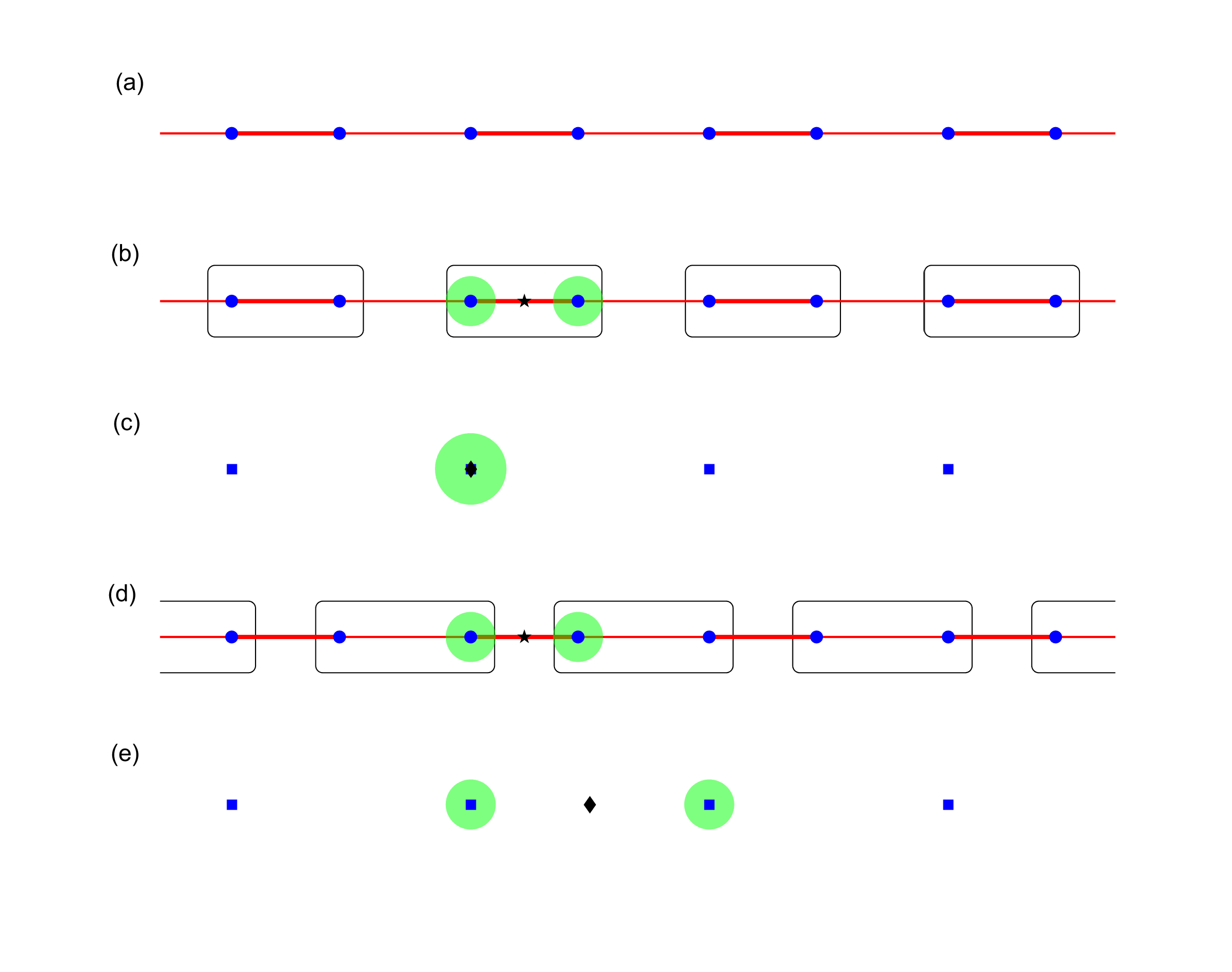}
    \caption{Same bulk SSH model with different choices of unit cells. (b) and (d) Two different choices of unit cells of the same bulk SSH model in (a). Orbits in the same rectangle are in the same unit cell. The probability distribution of the Wannier function is represented by the areas of the circle surrounding the orbits. (c) and (e) illustrate the probability distribution of the Wannier function if all the orbits in the same unit cell are assumed to be located at the lattice point. The black pentagrams show the actual positions of Wannier center. The black diamonds show the Wannier centers calculated from the Berry connection. }
    \label{fig1}
 \end{figure}
 In Fig.1 we illustrate that $x_w$ of SSH model depends on the choices of unit cells. We use the method in ref\cite{PhysRevB.96.245115} to construct the 1D Wannier function of the occupied bands of the SSH model. If we choose the unit cells in Fig.\ref{fig1}(b) $x_w$ will coincides with one lattice as is shown in Fig.\ref{fig1} (c) and the Zak phase $\phi_{Zak}=0$. If we  choose unit cells in Fig.\ref{fig1} (d) $x_w$ will be at the middle of two adjacent lattices as is shown in Fig.\ref{fig1} (e) and the Zak phase $\phi_{Zak}=\pi$.  The difference comes from the fact that the same orbit is considered to be located at different lattices with different choices of unit cells as is shown in Fig.\ref{fig1} (c) and (e).

It is clear the two values of Zak phase do not correspond to two different topological states of the bulk system, but only reflect two choices of unit cells of the same bulk system. As is shown in Fig.\ref{fig1} (b) and (d) the two choices of unit cells are equally possible. If two bulk SSH systems have different Zak phases, one can always change the choice of unit cells of one system and thus make the two systems have the same Zak phase. So there is only one bulk topological state of the SSH model when it is gapped. The so-called topological and so-called trivial state are only the same state with different types of unit cells.

\begin{figure}
  \includegraphics[width=9cm]{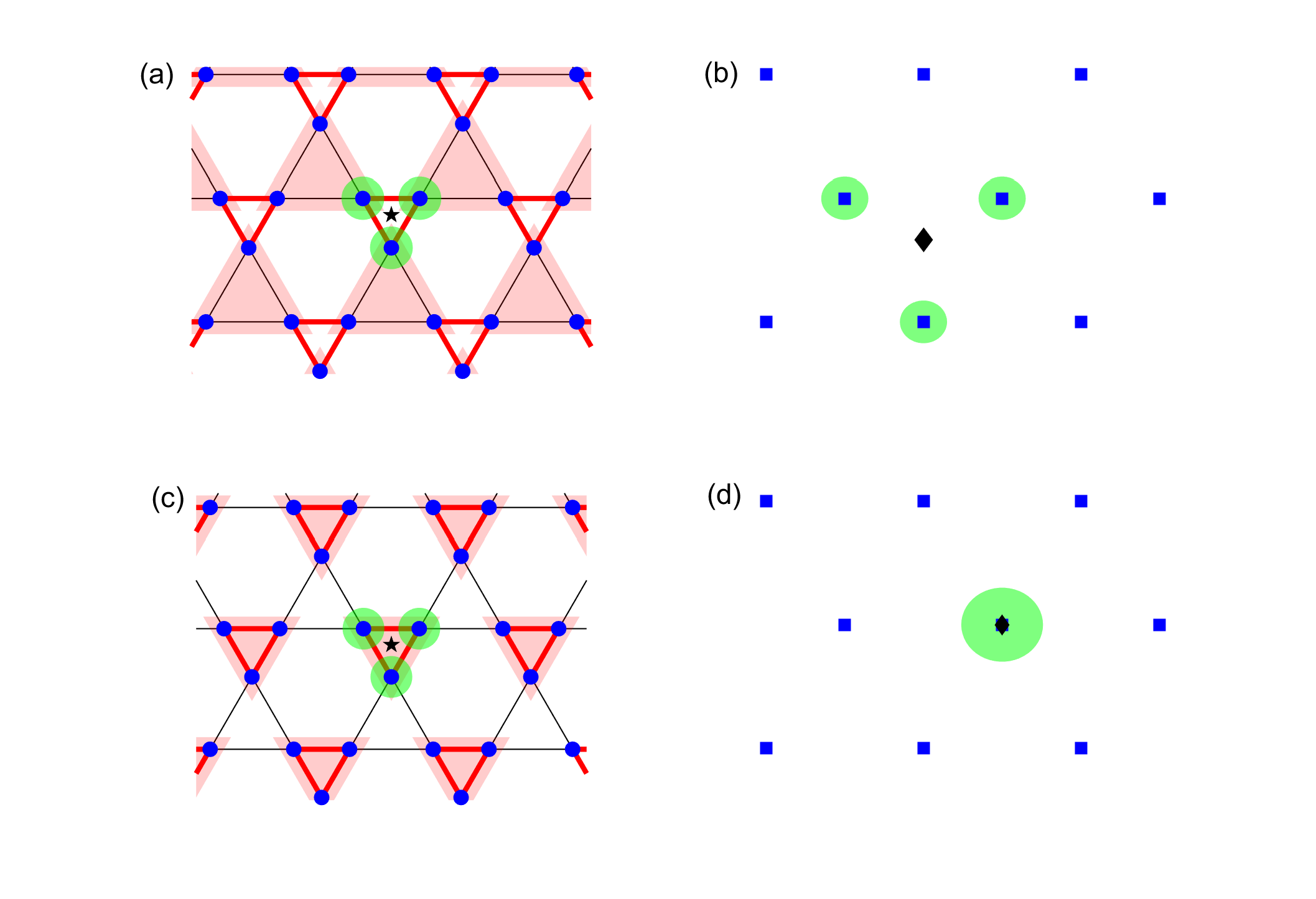}\label{kagome}
    \caption{(a) and (c) Same bulk breathing kagome lattice with different choices of unit cells. Orbits in the same red triangle are in the same unit cell. The probability distribution of the Wannier function is represented by the areas of the circle surrounding the orbits. (b) and (d) illustrate the probability distribution of the Wannier function if all the orbits in the same unit cell are assumed to be located at the lattice point. The black pentagrams show the actual position of Wannier center. The black diamonds show the Wannier centers calculated from the Berry connection.   }
    \label{fig2}
\end{figure}

\begin{figure}
  \includegraphics[width=9cm]{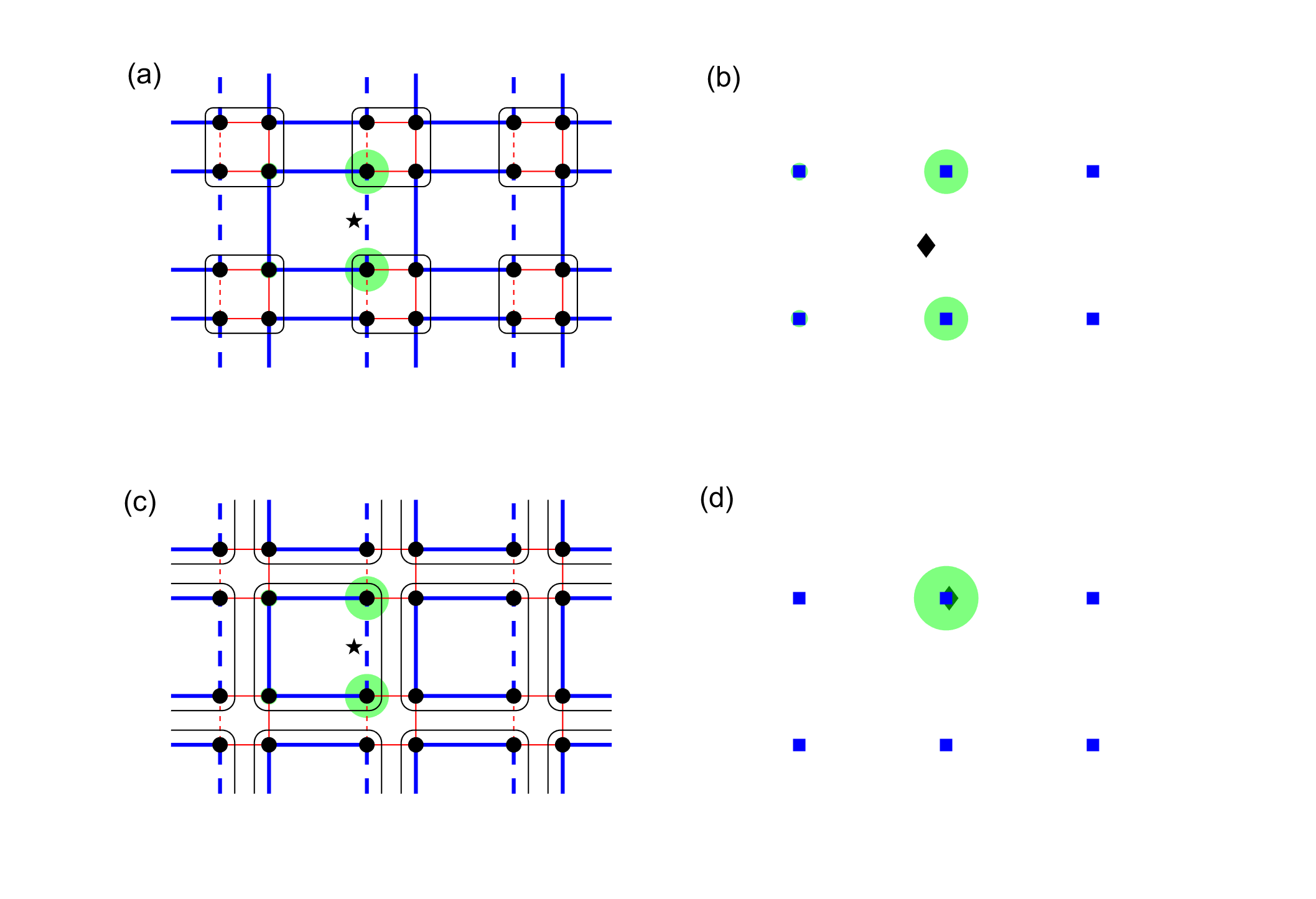}\label{quadrupole}
    \caption{(a) and (c) Same bulk system in ref\cite{Benalcazar61} with different choices of unit cells. Orbits in the same rectangle are in the same unit cell. The probability distribution of the Wannier function is represented by the areas of the circle surrounding the orbits. (b) and (d) illustrate the probability distribution of the Wannier function if all the orbits in the same unit cell are assumed to be located at the lattice point. The black pentagrams show the actual position of Wannier center. The black diamonds show the Wannier centers calculated from the Berry connection.  }
    \label{fig3}
\end{figure}

The Wannier centers $\mathbf{r}_w$ calculated from the Berry connections in 2D systems also depend on the choice of unit cells. In ref\cite{PhysRevLett.120.026801} it is claimed that the polarization is identical to the Wannier center. We can easily derive
  \begin{equation}\label{polarization}
\begin{split}
\mathbf{r}_w &=\frac{1}{S}\int_{BZ}\mathbf{A} d^2\mathbf{k}\\
             &=\sum\limits_{n}\mathbf{R}_n \langle W(\mathbf{R}_n)|W(\mathbf{R}_n)\rangle,
\end{split}
\end{equation}
where $\mathbf{A}=i\langle u (\mathbf{k})|\nabla_{\mathbf{k}}|u (\mathbf{k})\rangle$ is the Berry connection and $S$ is the area of the Brillouin zone. As in the 1D case $\langle W(\mathbf{R}_n)|W(\mathbf{R}_n)\rangle$ is the probability the electron in the unit cell around the lattice $\mathbf{R}_n$ when state of electron is described by the Wannier function $|W(\mathbf{R}_n)\rangle$. Clearly, all the orbits in the unit cell around the $\mathbf{R}_n$  are all assumed to be located at $\mathbf{R}_n$. So $\mathbf{r}_w$ is not the actual Wannier center.

In Fig.\ref{fig2} we show why $\mathbf{r}_w$ depends on the two equally possible choices of unit cells. We construct the 2D Wannier function by directly superposing the Bloch waves of occupied bands. Again, when we choose different unit cells the orbits will be consider to be located at different lattices. So the probability the electron is located at a lattice is different with different choices of unite cells as is shown in Fig.\ref{fig2} (b) and (d).

In ref\cite{Benalcazar61} the nested Wilson-loop is used to calculate the polarization of the Wannier band. Because the Chern number of the Wannier band is zero, localized Wannier functions can be constructed for the Wannier band. We superpose the Bloch waves of the Wannier band to obtain the Wannier function.  The polarization of Wannier band $p_y^{\nu_z^{\pm}}$ corresponds to the $y$ component of Wannier center calculated by \ref{polarization}.

Clearly, $p_y^{\nu_x^{\pm}}$ depends on the choices of unit cells. In Fig.\ref{fig3} we show two equally possible choices of unit cells. With the choice in Fig.\ref{fig3}(a) $p_y^{\nu_x^{-}}=\frac{1}{2}$ as is show in Fig.3(b). Use the same method we can obtain that $p_x^{\nu_y^{-}}=\frac{1}{2}$. So $(p_x^{\nu_y^{\pm}},p_y^{\nu_x^{\pm}})=(\frac{1}{2},\frac{1}{2})$ due to the reflection and inversion symmetries. Thus the so-called quadrupole invariant $q_{xy}=\frac{e}{2}$ and the bulk system will be considered as topological.

If we calculate the polarization with a different type unit cells for the same bulk system as shown in Fig.\ref{fig3} (c) and (d) we will find $(p_x^{\nu_y^{\pm}},p_y^{\nu_x^{\pm}})=(0,0)$ and $q_{xy}=0$. The system will be considered as trivial. Notice that the two types of unit cells have the same symmetry property, thus the two choices are equally possible. So we can not determine the so-called topological invariant $q_{xy}$ if we only look at the bulk system. Thus $q_{xy}$ is not a bulk topological invariant.

In conclusion the Wannier centers calculated from Berry connection with tight-binding models are not bulk topological invariants because they can not be uniquely defined. In fact, we should not attach much physical significance to such concepts because they only reflect the way we describe the system as is shown in Fig. 1-3.

Now we discuss the so-call topologically protected edge or corner states. In a finite system with boundary the nontrivial bulk topology manifest itself through the nontrivial boundary states. If the topology of bulk bands are protected by symmetry there are must be some symmetry constraint on the boundary, otherwise the boundary states may become trivial. For example, if the time reversal symmetry is broken at the boundary of a 2D or 3D topological insulator, the boundary states may become gapped.

We also stress some characteristics of the topological boundary states. First, the same bulk topological system must have the same boundary state property when we choose two types of boundaries with the same symmetry property. The two types of boundaries either both conform to the symmetry constraint or both violate the symmetry constraint. So the boundary states are either both topological or both trivial.

Second, since the existence of symmetry protected nontrivial boundary state is a distinctive feature of the topological system, such states should not exist in trivial systems no matter how we choose the boundary of the trivial system.
\begin{figure}
  \includegraphics[width=9cm]{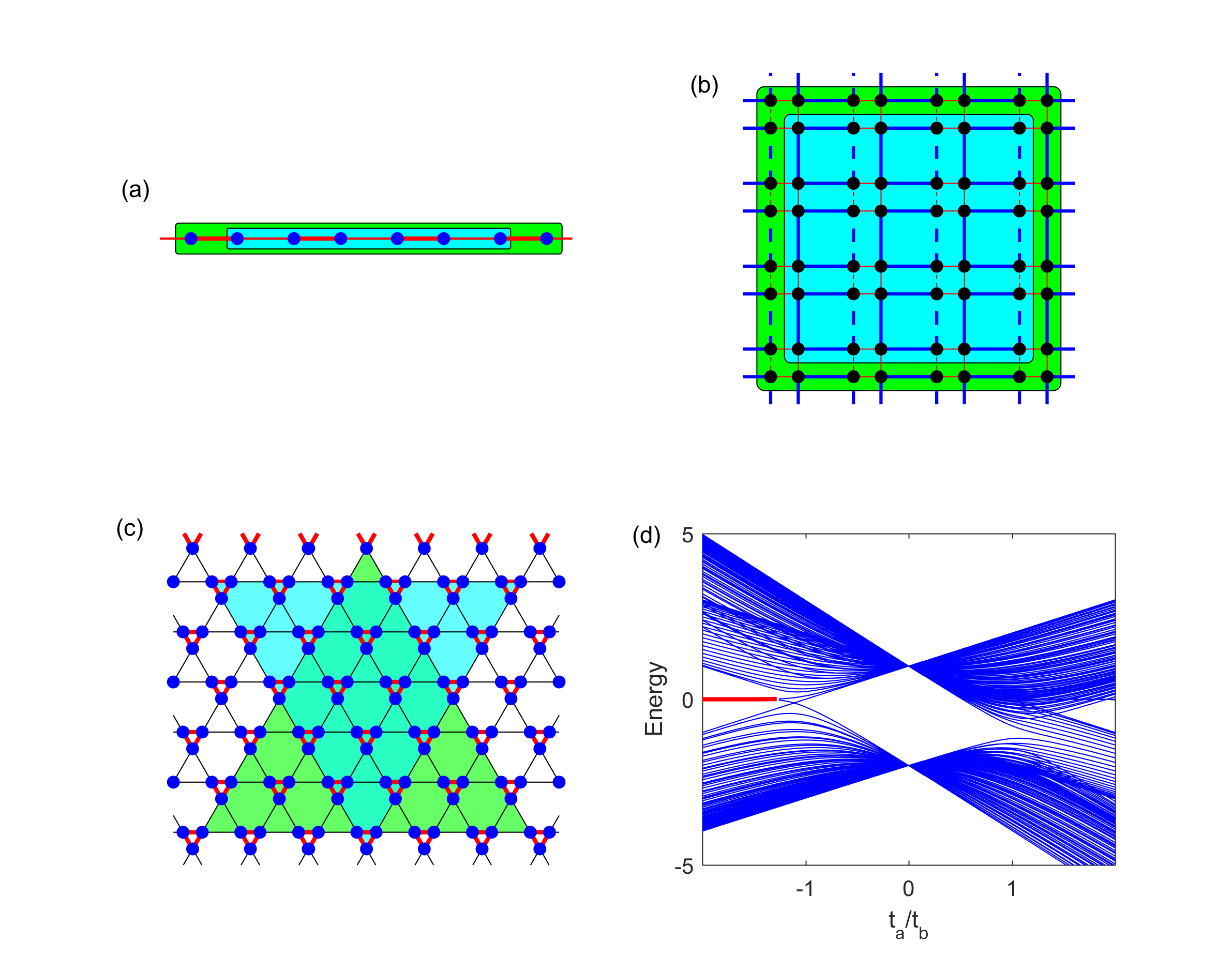}
    \caption{(a)-(c) Illustrations of same bulk system with different choices of boundaries. Open boundary condition is used for each boundary, i.e. the hoping amplitudes between the orbits inside and outside of the boundary are zero. (d)Energy spectrum of the $\nabla$ type of triangle in (c) with $L=20$.}
    \label{fig4}
\end{figure}

In Fig.\ref{fig4} (a)-(c) we show three systems, each with two types of boundaries. For each system the two types of boundaries have the same symmetry property. For example, in Fig.\ref{fig4}(a) the reflection symmetry and the chiral symmetry are both preserved with the two types of boundaries. So if the boundary state property is really topologically protected it must be the same for the two choices of boundaries. However, the boundary state properties are just opposite for the two boundaries: the zero-energy boundary states exist with one type of boundary and disappears with the other type.

In ref\cite{peterson2018a} they have not only observed the corner states but also claimed to test the robustness of the corner states by deforming the edge. However, as we show in Fig.\ref{fig4} (b), there are two types of boundaries with the same symmetry. If one consider the system is topological and the corner states are observed with one type of boundary, then the corner states can be obliterated by deforming boundary to the other type. More seriously, in the so-called trivial system we can always obtain equally robust corner states as the topological case by deforming the boundary. It is clearly impossible if the corner states are truly topologically protected.

If we choose the $\nabla$ type of boundary in Fig.\ref{fig4}(c) the zero-energy corner states disappear in the so-called topological phase and appear in the so-called trivial phase as is shown in Fig.\ref{fig4} (d). However, topological corner states should not exist in the trivial system no matter how we choose the boundary.

So the existence of the edge or corner states in these systems only reflect the choices of the boundaries of the same bulk system. One may argue if we choose one type of boundary then we must choose a commensurate type of unite cells, then the topological invariant of the bulk can be determined, and then the bulk system becomes topological or trivial. However, if the bulk topological property depends solely on the choice of the boundaries, is it still a bulk property?

If this kind of logic is valid, topological invariant of pure bulk system, i.e. infinite system or system with periodic boundary condition, will become meaningless because there is no boundary to determine the choice of unit cells.

In conclusion there is no topological distinction between the so-called topological phase and the so-called trivial phase in the systems discussed above. These systems only have one topological phase if the system is gapped. The so-called two topologically distinct phases are just the same bulk phase with different choices of unit cells. The existence or absence of the zero-energy boundary states only reflects the different choice of boundary of the same bulk systems.

One may contend that though the corner states do not reflect the bulk topology of the system, they do enjoy some topological protections, i.e. the two types of boundaries of the same bulk system discussed above can not be continuously connected without breaking the chiral symmetry. The measured corner states may be considered as topologically protected in this sense.  

First we must realize that this is just a boundary property, which is only concern with different choices boundaries for the same bulk system. It is does not mean that the bulk system have non-trivial and trivial states as is claimed by previous works.
  
Second this kind of systems can not be realized in quantum systems. The boundary properties depend on the chiral symmetry of the system. The chiral symmetry is a very artificial symmetry, which comes from the simplifications we make in constructing the tight-binding models. Strictly speaking no quantum system has chiral symmetry. For example, chiral symmetry requires that the energy spectrum of the model symmetrically distributed about zero. It is impossible because the energy spectrum of real quantum system is always lower bounded, but not upper bounded.
 
When we use tight-binding models and only consider a few bands, the chiral symmetry still requires unrealistic assumptions. The on-site potentials of all the orbits must be equal(Strictly speaking must be zero) and the next-nearest-neighbour hopping between the orbits must be forbidden. For simplicity open boundary conditions are used when we calculated the boundary properties of the tight-binding models. However, in real system there must be some difference between the on-site potentials of orbits inside of the system and those outside the system, i.e. the vacuum. So the chiral symmetry must be broken at the boundary of the real systems.

Even in the classical systems that are used to simulated the quantum systems the observed spectrums are far from being symmetric respect to the mid-gap point\cite{poli2015selective,peterson2018a,serragarcia2018observation,zhang2019second-order}. So the chiral symmetry is broken in these systems and the observed corner states are not protected even in this weak sense of topology protection.

\bibliographystyle{apsrev4-1}
\bibliography{references}

\begin{thebibliography}{24}%
\makeatletter
\providecommand \@ifxundefined [1]{%
 \@ifx{#1\undefined}
}%
\providecommand \@ifnum [1]{%
 \ifnum #1\expandafter \@firstoftwo
 \else \expandafter \@secondoftwo
 \fi
}%
\providecommand \@ifx [1]{%
 \ifx #1\expandafter \@firstoftwo
 \else \expandafter \@secondoftwo
 \fi
}%
\providecommand \natexlab [1]{#1}%
\providecommand \enquote  [1]{``#1''}%
\providecommand \bibnamefont  [1]{#1}%
\providecommand \bibfnamefont [1]{#1}%
\providecommand \citenamefont [1]{#1}%
\providecommand \href@noop [0]{\@secondoftwo}%
\providecommand \href [0]{\begingroup \@sanitize@url \@href}%
\providecommand \@href[1]{\@@startlink{#1}\@@href}%
\providecommand \@@href[1]{\endgroup#1\@@endlink}%
\providecommand \@sanitize@url [0]{\catcode `\\12\catcode `\$12\catcode
  `\&12\catcode `\#12\catcode `\^12\catcode `\_12\catcode `\%12\relax}%
\providecommand \@@startlink[1]{}%
\providecommand \@@endlink[0]{}%
\providecommand \url  [0]{\begingroup\@sanitize@url \@url }%
\providecommand \@url [1]{\endgroup\@href {#1}{\urlprefix }}%
\providecommand \urlprefix  [0]{URL }%
\providecommand \Eprint [0]{\href }%
\providecommand \doibase [0]{http://dx.doi.org/}%
\providecommand \selectlanguage [0]{\@gobble}%
\providecommand \bibinfo  [0]{\@secondoftwo}%
\providecommand \bibfield  [0]{\@secondoftwo}%
\providecommand \translation [1]{[#1]}%
\providecommand \BibitemOpen [0]{}%
\providecommand \bibitemStop [0]{}%
\providecommand \bibitemNoStop [0]{.\EOS\space}%
\providecommand \EOS [0]{\spacefactor3000\relax}%
\providecommand \BibitemShut  [1]{\csname bibitem#1\endcsname}%
\let\auto@bib@innerbib\@empty
\bibitem [{\citenamefont {Delplace}\ \emph {et~al.}(2011)\citenamefont
  {Delplace}, \citenamefont {Ullmo},\ and\ \citenamefont
  {Montambaux}}]{PhysRevB.84.195452}%
  \BibitemOpen
  \bibfield  {author} {\bibinfo {author} {\bibfnamefont {P.}~\bibnamefont
  {Delplace}}, \bibinfo {author} {\bibfnamefont {D.}~\bibnamefont {Ullmo}}, \
  and\ \bibinfo {author} {\bibfnamefont {G.}~\bibnamefont {Montambaux}},\
  }\href {\doibase 10.1103/PhysRevB.84.195452} {\bibfield  {journal} {\bibinfo
  {journal} {Phys. Rev. B}\ }\textbf {\bibinfo {volume} {84}},\ \bibinfo
  {pages} {195452} (\bibinfo {year} {2011})}\BibitemShut {NoStop}%
\bibitem [{\citenamefont {Ganeshan}\ \emph {et~al.}(2013)\citenamefont
  {Ganeshan}, \citenamefont {Sun},\ and\ \citenamefont
  {Das~Sarma}}]{PhysRevLett.110.180403}%
  \BibitemOpen
  \bibfield  {author} {\bibinfo {author} {\bibfnamefont {S.}~\bibnamefont
  {Ganeshan}}, \bibinfo {author} {\bibfnamefont {K.}~\bibnamefont {Sun}}, \
  and\ \bibinfo {author} {\bibfnamefont {S.}~\bibnamefont {Das~Sarma}},\ }\href
  {\doibase 10.1103/PhysRevLett.110.180403} {\bibfield  {journal} {\bibinfo
  {journal} {Phys. Rev. Lett.}\ }\textbf {\bibinfo {volume} {110}},\ \bibinfo
  {pages} {180403} (\bibinfo {year} {2013})}\BibitemShut {NoStop}%
\bibitem [{\citenamefont {Bahari}\ and\ \citenamefont
  {Hosseini}(2016)}]{PhysRevB.94.125119}%
  \BibitemOpen
  \bibfield  {author} {\bibinfo {author} {\bibfnamefont {M.}~\bibnamefont
  {Bahari}}\ and\ \bibinfo {author} {\bibfnamefont {M.~V.}\ \bibnamefont
  {Hosseini}},\ }\href {\doibase 10.1103/PhysRevB.94.125119} {\bibfield
  {journal} {\bibinfo  {journal} {Phys. Rev. B}\ }\textbf {\bibinfo {volume}
  {94}},\ \bibinfo {pages} {125119} (\bibinfo {year} {2016})}\BibitemShut
  {NoStop}%
\bibitem [{\citenamefont {Li}\ \emph {et~al.}(2014)\citenamefont {Li},
  \citenamefont {Xu},\ and\ \citenamefont {Chen}}]{PhysRevB.89.085111}%
  \BibitemOpen
  \bibfield  {author} {\bibinfo {author} {\bibfnamefont {L.}~\bibnamefont
  {Li}}, \bibinfo {author} {\bibfnamefont {Z.}~\bibnamefont {Xu}}, \ and\
  \bibinfo {author} {\bibfnamefont {S.}~\bibnamefont {Chen}},\ }\href {\doibase
  10.1103/PhysRevB.89.085111} {\bibfield  {journal} {\bibinfo  {journal} {Phys.
  Rev. B}\ }\textbf {\bibinfo {volume} {89}},\ \bibinfo {pages} {085111}
  (\bibinfo {year} {2014})}\BibitemShut {NoStop}%
\bibitem [{\citenamefont {Atala}\ \emph {et~al.}(2013)\citenamefont {Atala},
  \citenamefont {Aidelsburger}, \citenamefont {Barreiro}, \citenamefont
  {Abanin}, \citenamefont {Kitagawa}, \citenamefont {Demler},\ and\
  \citenamefont {Bloch}}]{atala2013direct}%
  \BibitemOpen
  \bibfield  {author} {\bibinfo {author} {\bibfnamefont {M.}~\bibnamefont
  {Atala}}, \bibinfo {author} {\bibfnamefont {M.}~\bibnamefont {Aidelsburger}},
  \bibinfo {author} {\bibfnamefont {J.~T.}\ \bibnamefont {Barreiro}}, \bibinfo
  {author} {\bibfnamefont {D.~A.}\ \bibnamefont {Abanin}}, \bibinfo {author}
  {\bibfnamefont {T.}~\bibnamefont {Kitagawa}}, \bibinfo {author}
  {\bibfnamefont {E.}~\bibnamefont {Demler}}, \ and\ \bibinfo {author}
  {\bibfnamefont {I.}~\bibnamefont {Bloch}},\ }\href@noop {} {\bibfield
  {journal} {\bibinfo  {journal} {Nature Physics}\ }\textbf {\bibinfo {volume}
  {9}},\ \bibinfo {pages} {795} (\bibinfo {year} {2013})}\BibitemShut {NoStop}%
\bibitem [{\citenamefont {Benalcazar}\ \emph
  {et~al.}(2017{\natexlab{a}})\citenamefont {Benalcazar}, \citenamefont
  {Bernevig},\ and\ \citenamefont {Hughes}}]{Benalcazar61}%
  \BibitemOpen
  \bibfield  {author} {\bibinfo {author} {\bibfnamefont {W.~A.}\ \bibnamefont
  {Benalcazar}}, \bibinfo {author} {\bibfnamefont {B.~A.}\ \bibnamefont
  {Bernevig}}, \ and\ \bibinfo {author} {\bibfnamefont {T.~L.}\ \bibnamefont
  {Hughes}},\ }\href {\doibase 10.1126/science.aah6442} {\bibfield  {journal}
  {\bibinfo  {journal} {Science}\ }\textbf {\bibinfo {volume} {357}},\ \bibinfo
  {pages} {61} (\bibinfo {year} {2017}{\natexlab{a}})},\ \Eprint
  {http://arxiv.org/abs/https://science.sciencemag.org/content/357/6346/61.full.pdf}
  {https://science.sciencemag.org/content/357/6346/61.full.pdf} \BibitemShut
  {NoStop}%
\bibitem [{\citenamefont {Benalcazar}\ \emph
  {et~al.}(2017{\natexlab{b}})\citenamefont {Benalcazar}, \citenamefont
  {Bernevig},\ and\ \citenamefont {Hughes}}]{PhysRevB.96.245115}%
  \BibitemOpen
  \bibfield  {author} {\bibinfo {author} {\bibfnamefont {W.~A.}\ \bibnamefont
  {Benalcazar}}, \bibinfo {author} {\bibfnamefont {B.~A.}\ \bibnamefont
  {Bernevig}}, \ and\ \bibinfo {author} {\bibfnamefont {T.~L.}\ \bibnamefont
  {Hughes}},\ }\href {\doibase 10.1103/PhysRevB.96.245115} {\bibfield
  {journal} {\bibinfo  {journal} {Phys. Rev. B}\ }\textbf {\bibinfo {volume}
  {96}},\ \bibinfo {pages} {245115} (\bibinfo {year}
  {2017}{\natexlab{b}})}\BibitemShut {NoStop}%
\bibitem [{\citenamefont {Imhof}\ \emph {et~al.}(2018)\citenamefont {Imhof},
  \citenamefont {Berger}, \citenamefont {Bayer}, \citenamefont {Brehm},
  \citenamefont {Molenkamp}, \citenamefont {Kiessling}, \citenamefont
  {Schindler}, \citenamefont {Lee}, \citenamefont {Greiter}, \citenamefont
  {Neupert} \emph {et~al.}}]{imhof2018topolectrical-circuit}%
  \BibitemOpen
  \bibfield  {author} {\bibinfo {author} {\bibfnamefont {S.}~\bibnamefont
  {Imhof}}, \bibinfo {author} {\bibfnamefont {C.}~\bibnamefont {Berger}},
  \bibinfo {author} {\bibfnamefont {F.}~\bibnamefont {Bayer}}, \bibinfo
  {author} {\bibfnamefont {J.}~\bibnamefont {Brehm}}, \bibinfo {author}
  {\bibfnamefont {L.~W.}\ \bibnamefont {Molenkamp}}, \bibinfo {author}
  {\bibfnamefont {T.}~\bibnamefont {Kiessling}}, \bibinfo {author}
  {\bibfnamefont {F.}~\bibnamefont {Schindler}}, \bibinfo {author}
  {\bibfnamefont {C.~H.}\ \bibnamefont {Lee}}, \bibinfo {author} {\bibfnamefont
  {M.}~\bibnamefont {Greiter}}, \bibinfo {author} {\bibfnamefont
  {T.}~\bibnamefont {Neupert}},  \emph {et~al.},\ }\href@noop {} {\bibfield
  {journal} {\bibinfo  {journal} {Nature Physics}\ }\textbf {\bibinfo {volume}
  {14}},\ \bibinfo {pages} {925} (\bibinfo {year} {2018})}\BibitemShut
  {NoStop}%
\bibitem [{\citenamefont {Serragarcia}\ \emph {et~al.}(2018)\citenamefont
  {Serragarcia}, \citenamefont {Peri}, \citenamefont {Susstrunk}, \citenamefont
  {Bilal}, \citenamefont {Larsen}, \citenamefont {Villanueva},\ and\
  \citenamefont {Huber}}]{serragarcia2018observation}%
  \BibitemOpen
  \bibfield  {author} {\bibinfo {author} {\bibfnamefont {M.}~\bibnamefont
  {Serragarcia}}, \bibinfo {author} {\bibfnamefont {V.}~\bibnamefont {Peri}},
  \bibinfo {author} {\bibfnamefont {R.}~\bibnamefont {Susstrunk}}, \bibinfo
  {author} {\bibfnamefont {O.~R.}\ \bibnamefont {Bilal}}, \bibinfo {author}
  {\bibfnamefont {T.}~\bibnamefont {Larsen}}, \bibinfo {author} {\bibfnamefont
  {L.~G.}\ \bibnamefont {Villanueva}}, \ and\ \bibinfo {author} {\bibfnamefont
  {S.~D.}\ \bibnamefont {Huber}},\ }\href@noop {} {\bibfield  {journal}
  {\bibinfo  {journal} {Nature}\ }\textbf {\bibinfo {volume} {555}},\ \bibinfo
  {pages} {342} (\bibinfo {year} {2018})}\BibitemShut {NoStop}%
\bibitem [{\citenamefont {Zhang}\ \emph {et~al.}(2019)\citenamefont {Zhang},
  \citenamefont {Wang}, \citenamefont {Lin}, \citenamefont {Tian},
  \citenamefont {Xie}, \citenamefont {Lu}, \citenamefont {Chen},\ and\
  \citenamefont {Jiang}}]{zhang2019second-order}%
  \BibitemOpen
  \bibfield  {author} {\bibinfo {author} {\bibfnamefont {X.}~\bibnamefont
  {Zhang}}, \bibinfo {author} {\bibfnamefont {H.}~\bibnamefont {Wang}},
  \bibinfo {author} {\bibfnamefont {Z.}~\bibnamefont {Lin}}, \bibinfo {author}
  {\bibfnamefont {Y.}~\bibnamefont {Tian}}, \bibinfo {author} {\bibfnamefont
  {B.}~\bibnamefont {Xie}}, \bibinfo {author} {\bibfnamefont {M.}~\bibnamefont
  {Lu}}, \bibinfo {author} {\bibfnamefont {Y.}~\bibnamefont {Chen}}, \ and\
  \bibinfo {author} {\bibfnamefont {J.}~\bibnamefont {Jiang}},\ }\href@noop {}
  {\bibfield  {journal} {\bibinfo  {journal} {Nature Physics}\ }\textbf
  {\bibinfo {volume} {15}},\ \bibinfo {pages} {582} (\bibinfo {year}
  {2019})}\BibitemShut {NoStop}%
\bibitem [{\citenamefont {Luo}\ and\ \citenamefont
  {Zhang}(2019)}]{PhysRevLett.123.073601}%
  \BibitemOpen
  \bibfield  {author} {\bibinfo {author} {\bibfnamefont {X.-W.}\ \bibnamefont
  {Luo}}\ and\ \bibinfo {author} {\bibfnamefont {C.}~\bibnamefont {Zhang}},\
  }\href {\doibase 10.1103/PhysRevLett.123.073601} {\bibfield  {journal}
  {\bibinfo  {journal} {Phys. Rev. Lett.}\ }\textbf {\bibinfo {volume} {123}},\
  \bibinfo {pages} {073601} (\bibinfo {year} {2019})}\BibitemShut {NoStop}%
\bibitem [{\citenamefont {Peterson}\ \emph {et~al.}(2018)\citenamefont
  {Peterson}, \citenamefont {Benalcazar}, \citenamefont {Hughes},\ and\
  \citenamefont {Bahl}}]{peterson2018a}%
  \BibitemOpen
  \bibfield  {author} {\bibinfo {author} {\bibfnamefont {C.~W.}\ \bibnamefont
  {Peterson}}, \bibinfo {author} {\bibfnamefont {W.~A.}\ \bibnamefont
  {Benalcazar}}, \bibinfo {author} {\bibfnamefont {T.~L.}\ \bibnamefont
  {Hughes}}, \ and\ \bibinfo {author} {\bibfnamefont {G.}~\bibnamefont
  {Bahl}},\ }\href@noop {} {\bibfield  {journal} {\bibinfo  {journal} {Nature}\
  }\textbf {\bibinfo {volume} {555}},\ \bibinfo {pages} {346} (\bibinfo {year}
  {2018})}\BibitemShut {NoStop}%
\bibitem [{\citenamefont {Ezawa}(2018{\natexlab{a}})}]{PhysRevLett.120.026801}%
  \BibitemOpen
  \bibfield  {author} {\bibinfo {author} {\bibfnamefont {M.}~\bibnamefont
  {Ezawa}},\ }\href {\doibase 10.1103/PhysRevLett.120.026801} {\bibfield
  {journal} {\bibinfo  {journal} {Phys. Rev. Lett.}\ }\textbf {\bibinfo
  {volume} {120}},\ \bibinfo {pages} {026801} (\bibinfo {year}
  {2018}{\natexlab{a}})}\BibitemShut {NoStop}%
\bibitem [{\citenamefont {Ezawa}(2018{\natexlab{b}})}]{PhysRevB.98.045125}%
  \BibitemOpen
  \bibfield  {author} {\bibinfo {author} {\bibfnamefont {M.}~\bibnamefont
  {Ezawa}},\ }\href {\doibase 10.1103/PhysRevB.98.045125} {\bibfield  {journal}
  {\bibinfo  {journal} {Phys. Rev. B}\ }\textbf {\bibinfo {volume} {98}},\
  \bibinfo {pages} {045125} (\bibinfo {year} {2018}{\natexlab{b}})}\BibitemShut
  {NoStop}%
\bibitem [{\citenamefont {Kempkes}\ \emph {et~al.}(2019)\citenamefont
  {Kempkes}, \citenamefont {Slot}, \citenamefont {Den~Broeke}, \citenamefont
  {Capiod}, \citenamefont {Benalcazar}, \citenamefont {Vanmaekelbergh},
  \citenamefont {Bercioux}, \citenamefont {Swart},\ and\ \citenamefont
  {Smith}}]{kempkes2019robust}%
  \BibitemOpen
  \bibfield  {author} {\bibinfo {author} {\bibfnamefont {S.~N.}\ \bibnamefont
  {Kempkes}}, \bibinfo {author} {\bibfnamefont {M.~R.}\ \bibnamefont {Slot}},
  \bibinfo {author} {\bibfnamefont {J.~J.~V.}\ \bibnamefont {Den~Broeke}},
  \bibinfo {author} {\bibfnamefont {P.}~\bibnamefont {Capiod}}, \bibinfo
  {author} {\bibfnamefont {W.~A.}\ \bibnamefont {Benalcazar}}, \bibinfo
  {author} {\bibfnamefont {D.}~\bibnamefont {Vanmaekelbergh}}, \bibinfo
  {author} {\bibfnamefont {D.}~\bibnamefont {Bercioux}}, \bibinfo {author}
  {\bibfnamefont {I.}~\bibnamefont {Swart}}, \ and\ \bibinfo {author}
  {\bibfnamefont {C.~M.}\ \bibnamefont {Smith}},\ }\href@noop {} {\bibfield
  {journal} {\bibinfo  {journal} {Nature Materials}\ ,\ \bibinfo {pages} {1}}
  (\bibinfo {year} {2019})}\BibitemShut {NoStop}%
\bibitem [{\citenamefont {Ni}\ \emph {et~al.}(2019)\citenamefont {Ni},
  \citenamefont {Weiner}, \citenamefont {Alu},\ and\ \citenamefont
  {Khanikaev}}]{ni2019observation}%
  \BibitemOpen
  \bibfield  {author} {\bibinfo {author} {\bibfnamefont {X.}~\bibnamefont
  {Ni}}, \bibinfo {author} {\bibfnamefont {M.}~\bibnamefont {Weiner}}, \bibinfo
  {author} {\bibfnamefont {A.}~\bibnamefont {Alu}}, \ and\ \bibinfo {author}
  {\bibfnamefont {A.~B.}\ \bibnamefont {Khanikaev}},\ }\href@noop {} {\bibfield
   {journal} {\bibinfo  {journal} {Nature Materials}\ }\textbf {\bibinfo
  {volume} {18}},\ \bibinfo {pages} {113} (\bibinfo {year} {2019})}\BibitemShut
  {NoStop}%
\bibitem [{\citenamefont {Mittal}\ \emph {et~al.}(2019)\citenamefont {Mittal},
  \citenamefont {Orre}, \citenamefont {Zhu}, \citenamefont {Gorlach},
  \citenamefont {Poddubny},\ and\ \citenamefont {Hafezi}}]{mittal2019photonic}%
  \BibitemOpen
  \bibfield  {author} {\bibinfo {author} {\bibfnamefont {S.}~\bibnamefont
  {Mittal}}, \bibinfo {author} {\bibfnamefont {V.~V.}\ \bibnamefont {Orre}},
  \bibinfo {author} {\bibfnamefont {G.}~\bibnamefont {Zhu}}, \bibinfo {author}
  {\bibfnamefont {M.~A.}\ \bibnamefont {Gorlach}}, \bibinfo {author}
  {\bibfnamefont {A.~N.}\ \bibnamefont {Poddubny}}, \ and\ \bibinfo {author}
  {\bibfnamefont {M.}~\bibnamefont {Hafezi}},\ }\href@noop {} {\bibfield
  {journal} {\bibinfo  {journal} {Nature Photonics}\ }\textbf {\bibinfo
  {volume} {13}},\ \bibinfo {pages} {692} (\bibinfo {year} {2019})}\BibitemShut
  {NoStop}%
\bibitem [{\citenamefont {Hassan}\ \emph {et~al.}(2019)\citenamefont {Hassan},
  \citenamefont {Kunst}, \citenamefont {Moritz}, \citenamefont {Andler},
  \citenamefont {Bergholtz},\ and\ \citenamefont
  {Bourennane}}]{hassan2019corner}%
  \BibitemOpen
  \bibfield  {author} {\bibinfo {author} {\bibfnamefont {A.~E.}\ \bibnamefont
  {Hassan}}, \bibinfo {author} {\bibfnamefont {F.~K.}\ \bibnamefont {Kunst}},
  \bibinfo {author} {\bibfnamefont {A.}~\bibnamefont {Moritz}}, \bibinfo
  {author} {\bibfnamefont {G.}~\bibnamefont {Andler}}, \bibinfo {author}
  {\bibfnamefont {E.~J.}\ \bibnamefont {Bergholtz}}, \ and\ \bibinfo {author}
  {\bibfnamefont {M.}~\bibnamefont {Bourennane}},\ }\href@noop {} {\bibfield
  {journal} {\bibinfo  {journal} {Nature Photonics}\ }\textbf {\bibinfo
  {volume} {13}},\ \bibinfo {pages} {697} (\bibinfo {year} {2019})}\BibitemShut
  {NoStop}%
\bibitem [{\citenamefont {Xue}\ \emph {et~al.}(2019)\citenamefont {Xue},
  \citenamefont {Yang}, \citenamefont {Liu}, \citenamefont {Gao}, \citenamefont
  {Chong},\ and\ \citenamefont {Zhang}}]{PhysRevLett.122.244301}%
  \BibitemOpen
  \bibfield  {author} {\bibinfo {author} {\bibfnamefont {H.}~\bibnamefont
  {Xue}}, \bibinfo {author} {\bibfnamefont {Y.}~\bibnamefont {Yang}}, \bibinfo
  {author} {\bibfnamefont {G.}~\bibnamefont {Liu}}, \bibinfo {author}
  {\bibfnamefont {F.}~\bibnamefont {Gao}}, \bibinfo {author} {\bibfnamefont
  {Y.}~\bibnamefont {Chong}}, \ and\ \bibinfo {author} {\bibfnamefont
  {B.}~\bibnamefont {Zhang}},\ }\href {\doibase 10.1103/PhysRevLett.122.244301}
  {\bibfield  {journal} {\bibinfo  {journal} {Phys. Rev. Lett.}\ }\textbf
  {\bibinfo {volume} {122}},\ \bibinfo {pages} {244301} (\bibinfo {year}
  {2019})}\BibitemShut {NoStop}%
\bibitem [{\citenamefont {Benalcazar}\ \emph {et~al.}(2019)\citenamefont
  {Benalcazar}, \citenamefont {Li},\ and\ \citenamefont
  {Hughes}}]{PhysRevB.99.245151}%
  \BibitemOpen
  \bibfield  {author} {\bibinfo {author} {\bibfnamefont {W.~A.}\ \bibnamefont
  {Benalcazar}}, \bibinfo {author} {\bibfnamefont {T.}~\bibnamefont {Li}}, \
  and\ \bibinfo {author} {\bibfnamefont {T.~L.}\ \bibnamefont {Hughes}},\
  }\href {\doibase 10.1103/PhysRevB.99.245151} {\bibfield  {journal} {\bibinfo
  {journal} {Phys. Rev. B}\ }\textbf {\bibinfo {volume} {99}},\ \bibinfo
  {pages} {245151} (\bibinfo {year} {2019})}\BibitemShut {NoStop}%
\bibitem [{\citenamefont {Chen}\ \emph {et~al.}(2019)\citenamefont {Chen},
  \citenamefont {Xu}, \citenamefont {Al~Jahdali}, \citenamefont {Mei},\ and\
  \citenamefont {Wu}}]{PhysRevB.100.075120}%
  \BibitemOpen
  \bibfield  {author} {\bibinfo {author} {\bibfnamefont {Z.-G.}\ \bibnamefont
  {Chen}}, \bibinfo {author} {\bibfnamefont {C.}~\bibnamefont {Xu}}, \bibinfo
  {author} {\bibfnamefont {R.}~\bibnamefont {Al~Jahdali}}, \bibinfo {author}
  {\bibfnamefont {J.}~\bibnamefont {Mei}}, \ and\ \bibinfo {author}
  {\bibfnamefont {Y.}~\bibnamefont {Wu}},\ }\href {\doibase
  10.1103/PhysRevB.100.075120} {\bibfield  {journal} {\bibinfo  {journal}
  {Phys. Rev. B}\ }\textbf {\bibinfo {volume} {100}},\ \bibinfo {pages}
  {075120} (\bibinfo {year} {2019})}\BibitemShut {NoStop}%
\bibitem [{\citenamefont {Poddubny}(2019)}]{PhysRevB.100.075418}%
  \BibitemOpen
  \bibfield  {author} {\bibinfo {author} {\bibfnamefont {A.~N.}\ \bibnamefont
  {Poddubny}},\ }\href {\doibase 10.1103/PhysRevB.100.075418} {\bibfield
  {journal} {\bibinfo  {journal} {Phys. Rev. B}\ }\textbf {\bibinfo {volume}
  {100}},\ \bibinfo {pages} {075418} (\bibinfo {year} {2019})}\BibitemShut
  {NoStop}%
\bibitem [{\citenamefont {Pelegr\'{\i}}\ \emph {et~al.}(2019)\citenamefont
  {Pelegr\'{\i}}, \citenamefont {Marques}, \citenamefont {Ahufinger},
  \citenamefont {Mompart},\ and\ \citenamefont {Dias}}]{PhysRevB.100.205109}%
  \BibitemOpen
  \bibfield  {author} {\bibinfo {author} {\bibfnamefont {G.}~\bibnamefont
  {Pelegr\'{\i}}}, \bibinfo {author} {\bibfnamefont {A.~M.}\ \bibnamefont
  {Marques}}, \bibinfo {author} {\bibfnamefont {V.}~\bibnamefont {Ahufinger}},
  \bibinfo {author} {\bibfnamefont {J.}~\bibnamefont {Mompart}}, \ and\
  \bibinfo {author} {\bibfnamefont {R.~G.}\ \bibnamefont {Dias}},\ }\href
  {\doibase 10.1103/PhysRevB.100.205109} {\bibfield  {journal} {\bibinfo
  {journal} {Phys. Rev. B}\ }\textbf {\bibinfo {volume} {100}},\ \bibinfo
  {pages} {205109} (\bibinfo {year} {2019})}\BibitemShut {NoStop}%
\bibitem [{\citenamefont {Poli}\ \emph {et~al.}(2015)\citenamefont {Poli},
  \citenamefont {Bellec}, \citenamefont {Kuhl}, \citenamefont {Mortessagne},\
  and\ \citenamefont {Schomerus}}]{poli2015selective}%
  \BibitemOpen
  \bibfield  {author} {\bibinfo {author} {\bibfnamefont {C.}~\bibnamefont
  {Poli}}, \bibinfo {author} {\bibfnamefont {M.}~\bibnamefont {Bellec}},
  \bibinfo {author} {\bibfnamefont {U.}~\bibnamefont {Kuhl}}, \bibinfo {author}
  {\bibfnamefont {F.}~\bibnamefont {Mortessagne}}, \ and\ \bibinfo {author}
  {\bibfnamefont {H.}~\bibnamefont {Schomerus}},\ }\href@noop {} {\bibfield
  {journal} {\bibinfo  {journal} {Nature Communications}\ }\textbf {\bibinfo
  {volume} {6}},\ \bibinfo {pages} {6710} (\bibinfo {year} {2015})}\BibitemShut
  {NoStop}%
\end{thebibliography}%
\end{document}